\documentclass[final,12pt]{article}

\usepackage[pagebackref,colorlinks,citecolor=blue]{hyperref}
\usepackage{amsfonts}
\usepackage{amsmath}
\usepackage{times}
\usepackage{amssymb}
\usepackage{enumerate} 
\usepackage{natbib} 
\usepackage{graphicx}
\usepackage[dvipsnames]{xcolor}
\usepackage[margin=3cm]{geometry}

\newcommand{\red}[1]{\textcolor{red}{#1}}
\newcommand{\blue}[1]{\textcolor{blue}{#1}}
\newcommand{\purple}[1]{\textcolor{Purple}{#1}}

\newcommand{\xbf}{\boldsymbol{x}}
\newcommand{\ybf}{\boldsymbol{y}}
\newcommand{\gammabf}{\boldsymbol{\gamma}}
\newcommand{\Ycf}{\mathcal{Y}}

\newcommand{\predsetj}{A_\alpha(\xbf_j)}
\newcommand{\directpredsetj}{A^{D}_\alpha(\Xbf_j)}

\newcommand{\optpredsetj}{A^{I}_\alpha(\Xbf_j)}

\newcommand{\orderedpredset}{A_\alpha^{\theta,o}}

\newtheorem{theorem}{Theorem}

\newtheorem{definition}{Definition}

\graphicspath{{./plots/}}

\title{Frequentist Prediction Sets for Species Abundance using Indirect Information 
} 
\author{Elizabeth Bersson and Peter D. Hoff \\
Department of Statistical Science\\
Duke University}

\usepackage{bbm}

\newcommand\one{\mathbf{1}}

\newcommand\thetabf{{\boldsymbol{\theta}}}
\newcommand\Ybf{\boldsymbol{Y}}
\newcommand\Xbf{\boldsymbol{X}}
\newcommand\obf{\boldsymbol{o}}

\begin{document}

\maketitle

\abstract{
Citizen science databases that consist of volunteer-led sampling efforts of species
communities 
are relied on 
as essential sources of data
in ecology.
Summarizing such data across counties with frequentist-valid
prediction sets for each county
provides 
an interpretable comparison across counties of varying size or composition.
As citizen science data often feature unequal sampling efforts across a spatial domain,
prediction sets constructed with indirect methods that share information across counties may be used to improve precision.
In this article, we 
present a nonparametric framework
to obtain
precise prediction sets for a multinomial random sample based on indirect information that maintain frequentist coverage guarantees for each county.
We detail a simple algorithm 
to obtain prediction sets for each county using indirect information 
where the computation time does not depend on the sample size and scales nicely with the number of species considered.
The indirect information may be estimated by a proposed empirical Bayes procedure based on information from auxiliary data.
Our approach makes inference for under-sampled counties more precise, while maintaining 
area-specific frequentist validity for each county. 
Our method is used to provide a useful description of avian species abundance in North Carolina, USA
based on citizen science data from the eBird database.

}


\section{Introduction}

Understanding species abundance 
across heterogeneous spatial areas is an important task in ecology.
Citizen science databases that consist of observations of species counts gathered by volunteers are increasingly regarded as one of the richest sources of data for such a task.
One of the largest such data sources is the
eBird database
in which citizen scientists
throughout the world 
input 
counts of bird sightings \citep{Sullivan2009}.
In addition to 
its use for
describing avian species abundance,
eBird is a principal resource for understanding global biodiversity and is widely used in constructing and implementing conservation action plans \citep{Sullivan2017}.


More generally,
analyses from such databases may be used for informing policy, conservation efforts, habitat preservation, and more, for which understanding species prevalence
for non-overlapping geographic areas, such as counties across a state or country, 
is important. 
In practice, species abundance from citizen science data are commonly summarised within areas such as counties by empirical proportions from a sample, as in, e.g., \cite{Arnold2021,Camerini2014}.
Such proportions can be used to construct a prediction set for each county
that provides a description of species prevalence for that county with guaranteed frequentist coverage.


Given the impact on policy design,
corresponding uncertainty quantification is of particular import \citep{Lele2020a},
and so
it is desirable that precise prediction sets maintain a target coverage rate regardless of the county's size or composition.
This is challenging as a common feature of citizen science data is unequal sampling efforts that results in some counties with large amounts of data information and others with very little.
Using direct procedures that only make use of within-county information, a prediction set 
may be imprecise
in these counties with low sampling efforts.
This suggests using indirect information such as data from neighboring counties to improve prediction set precision for a given county.

In this article, we 
describe species abundance across sampling areas such as counties
with frequentist-valid prediction sets that are constructed to contain an unobserved bird with $1-\alpha$ probability. 
That is, a valid prediction set for a given county 
is a set of avian species such that an unobserved bird will belong to one of those species with $1-\alpha$ probability in a frequentist sense.
We develop a valid
nonparametric prediction method that allows for information to be shared across counties. 
Specifically, our approach
results in prediction sets with guaranteed frequentist coverage for each county that are constructed with the incorporation of indirect or prior information.
We detail and provide code for an empirical Bayes procedure to estimate such prior information from auxiliary data such as neighboring counties.
If this indirect information used to construct the prediction sets is accurate, the prediction sets will be smaller than 
direct sets that only make use of within-county information.

In Section \ref{sec4}, we detail the usefulness of the proposed approach in summarising the eBird citizen science data. 
Sharing information across counties generally results in smaller prediction sets as compared to direct prediction approaches, particularly so in counties with low sampling efforts. 
Moreover, the prediction sets provide a useful summary of the data that may be used to compare information across areas and better inform policy.

\section{Methodology}\label{sec2}

\subsection{Background and Notation}\label{subsec1}

For county $j\in\{1,...,J\}$, let $\Xbf_j$ be a vector of length $K$ 
where $X_{j,i}=x_{j,i}$ is the observed count of species $i$ over some set sampling period that may vary across counties.
We model $\Xbf_j$ with a $K$-dimensional multinomial distribution with $N_j=\sum_{i=1}^K x_{j,i}$ trials
and population proportions vector $\thetabf_j$,
\begin{equation}\label{samplingmodel}
\Xbf_j \sim MN_K(\thetabf_j,N_j).
\end{equation}
We construct a prediction set $\predsetj$ for 
an observation of a new bird arising from the same distribution, $\Ybf_j\sim MN_K(\thetabf_j,1)$ where $\Ybf_j\in\Ycf$ 
for $\Ycf = \{(y_1,...,y_K):\sum_{i=1}^Ky_i = 1,y_i\in\{0,1\}(i=1,....,K)\}$.
Let $\ybf_j^{(k)}\in\mathcal{Y}$ denote a prediction of category $k$, that is, let $\ybf_j^{(k)}$ be a vector of length $K$ with a one at index $k$ and zeros elsewhere.
In particular, we are interested in a prediction set for $\Ybf_j$
that maintains frequentist validity for some error rate $\alpha$.
Formally, we refer to this as an $\alpha$-valid prediction set:
    \begin{definition}[$\alpha$-Valid Prediction Set]
    An $\alpha$-valid prediction set for a predictand $\Ybf_j\in\mathcal{Y}$ is any subset $A_\alpha$ of the sample space $\mathcal{Y}$ that contains $\Ybf_j$ with probability greater than or equal to $1-\alpha$,
    \begin{equation}\label{validpred}
    P_\theta\left(\Ybf_j\in \predsetj\right) \geq 1-\alpha,\quad \forall\;\theta,
    \end{equation}
    where the probability is taken with respect to $\Ybf_j$ and $\Xbf_j$.
    \end{definition}
Additionally, small or precise $\alpha$-valid prediction sets are of particular interest, where prediction set size is measured by expected cardinality, that is, expected number of the $K$ categories in the sample space included in the prediction set.

\subsection{Order-based prediction for a single area}

A standard approach to construct $\alpha$-valid prediction sets for each county or area
is with a
direct method that only makes use of within-area information.
As such, we first consider construction of a prediction set for a single area $j$, using only data from county $j$. For ease of notation, we drop the area-identifying subscript in this subsection.

For multinomial data in general, if the event probability vector $\thetabf$ is known, an $\alpha$-valid prediction set is any combination of categories such that their event probabilities cumulatively sum to be greater than or equal to $1-\alpha$.
Equivalently stated, an $\alpha$-valid prediction set may be constructed by excluding categories such that the cumulative sum of the excluded categories' event probabilities is less than $\alpha$.
Such a prediction set may be constructed by admitting categories in some prespecified order into the prediction set until the cumulative sum of their event probabilities is at least $1-\alpha$.
The resulting prediction set will have $1-\alpha$ coverage regardless of the ordering used to admit categories.
In fact, the class of all $\alpha$-valid prediction sets may be constructed by following this procedure for non-strict total orderings of categories.

Perhaps intuitively, 
constructing such a prediction set by including categories with the largest event probabilities will result in the smallest $\alpha$-valid prediction set. 
In the terminology of ordering, this corresponds to 
constructing a prediction set based on
an ordering of categories that matches the ordering of the elements in $\thetabf$. 
We refer to this optimal ordering as the oracle ordering:

\begin{theorem}[Oracle order-based prediction]\label{thm1}
    Let $\Ybf\sim MN_K(\thetabf,1)$ for $\thetabf$ known. Then,
    \begin{enumerate}[(a)]
    \item the class of all $\alpha$-valid prediction sets for a given $\thetabf$
    consists of prediction sets of the form,
    \begin{equation}
    \orderedpredset = \left\{\ybf^{(k)} \in\Ycf :\left[ \sum_{l=1}^K \one \left(o_{k}\geq o_{l}\right) \theta_{l}\right] > \alpha\right\}\label{knowntheta_orderedpred},
    \end{equation}
    for some vector $\obf\in \mathbb{R}^K$, and
    \item the \textbf{oracle ordering} is that which corresponds to the increasing order statistics of $\thetabf$,
    \[
    \obf^{\theta} = \left\{\obf : 
    \theta_{m}<\theta_{n} \Rightarrow o_{m}<o_{n} \;\forall \;m,n\in\{1,....,K\}, m\ne n
    \right\},
    \]
     and $A_\alpha^{\theta,o^{\theta}}$ has the smallest cardinality among all orderings. 
    \end{enumerate}
\end{theorem}


In practice, $\thetabf$ is unknown, but a prediction set may be constructed based on an observed sample $\Xbf=\xbf$. 
It turns out, in fact, that any conditional $\alpha$-valid prediction set can be written 
similarly to the previous
construction (Equation \ref{knowntheta_orderedpred})
where the cumulative sum is computed with respect to the empirical proportions 
given by $\xbf$ and $\ybf$. 
This is a generalization of the conformal prediction framework, a popular machine learning approach to construct prediction regions based on measuring conformity (or non-conformity) of a predictand to an observed sample \citep{Vovk2005}.

\begin{theorem}[$\alpha$-valid order-based prediction]\label{allconformal}
Let $\Xbf\sim MN_K(\thetabf,N),\Ybf\sim MN_K(\thetabf,1)$. Then, every conformal $\alpha$-valid prediction set based on observed data $\xbf$ can be written
\begin{equation}\label{unknownprob_predset}
A_\alpha(\xbf) = \left\{ \ybf^{(k)}\in\mathcal{Y} : \left[\sum_{l=1}^K \mathbbm{1}\left(o_k\geq o_l\right) \frac{x_l+y_l^{(k)}}{N+1}\right]>\alpha\right\},
\end{equation}
for some vector $\obf\in \mathbb{R}^K$.
\end{theorem}
\noindent Note that the prediction set depends on the vector $\obf$ only through the order of its elements.


For any ordering of the $K$ categories, constructing a prediction set following Theorem \ref{allconformal} results in a prediction set with guaranteed finite-sample $1-\alpha$ frequentist coverage. 
The choice of ordering, however, will impact prediction set precision, that is, the set's cardinality.
For inference for a single area, 
a natural approach is to order the categories 
with respect to
their empirical proportions. 
The empirical proportions are unbiased for population proportions, so, if the area has a large sample size, an ordering based on the empirical proportions will approximate the oracle ordering well.
It turns out this approach is well-motivated by classical prediction approaches. 
Specifically, a standard
direct prediction method constructs a prediction set separately for an area based on an area-specific conditional pivotal quantity \citep{Faulkenberry1973,Tian2022a}. 
For a multinomial population, $\Ybf|\Xbf+\Ybf$ is such a quantity that follows a multivariate hypergeometric distribution which does not depend on the event probability vector. 
See \cite{Thatcher1964} for work on prediction sets of this type for binomial data. 
A prediction set constructed to contain species belonging to a highest mass region of this pivotal distribution is obtained by including species with the largest empirical counts until their cumulative proportion sum exceeds $1-\alpha$,
\begin{align}\label{directpredeqn}
    A_\alpha^D {}& (\xbf) =  
    \Big\{ \ybf^{(k)}\in\mathcal{Y} : \\
    {}& \left[\sum_{l=1}^K \mathbbm{1}\left(\left(x_{k} + y_k^{(k)}\right) \geq \left(x_{l}+y^{(k)}_l\right)\right) \frac{x_{l}+y_l^{(k)}}{N+1}\right]>\alpha\Big\}\nonumber.
\end{align}

This direct prediction set based on an ordering of the empirical proportions
is appealing as it is easy to interpret and
has finite-sample guaranteed $1-\alpha$ frequentist coverage.
For an area with low sampling effort, though, 
the empirical proportions will not precisely estimate the true proportions. As a result, 
a prediction set may have prohibitively large cardinality such that it is not practically useful. 
For such an area, incorporating indirect information from neighboring counties can 
improve the estimates of the county proportions and thereby increase the precision of
a prediction set.

\subsection{Order-based prediction for multiple areas}


In general, in analyzing
small area data, that is, 
areal data featuring small within-area sample sizes in some areas,
it is common to utilize indirect methods that share information across areas \citep{Rao2015}. 
The eBird database is a rich data source, and inference in any given county may be improved upon by taking advantage of auxiliary data using an indirect method.
In this subsection, we detail how information from neighboring counties may be used in 
estimating an ordering of categories
to improve prediction set precision.

As opposed to a direct prediction set based on an ordering corresponding to within-county empirical proportions,
an indirect prediction set can be constructed similarly whereby species are admitted into the prediction set based on an ordering corresponding to empirical posterior proportions estimated from a hierarchical model.
Such an estimate may be obtained based on a
conjugate Dirichlet prior distribution 
parameterized 
with a
common concentration hyperparameter
for the $J$ areas, 
\begin{equation}\label{linkingmodel}
\thetabf_1,...,\thetabf_J\sim {Dirichlet}_K(\boldsymbol{\gamma}).
\end{equation}
Given a hyperparameter $\gammabf\in\mathbb{R}^K$,
the posterior 
expectation of the proportions $\thetabf_j$ in county $j$ is 
$\tilde{\xbf}_{j}/(N_j + \sum_{i=1}^K \gamma_i)$ where $\tilde{\xbf}_{j} = \xbf_j+\gammabf$.
In this way, $\tilde{\xbf}_{j}$ may be interpreted as a posterior vector of counts for county $j$.
Then, 
an $\alpha$-valid prediction set based on $\tilde{\xbf}_{j}$ is,
\begin{align}\label{indirectpredset}
    A_\alpha^I {}& (\xbf_j) =
    \Big\{ \ybf^{(k)}\in\mathcal{Y} : \\
    {}& \left[\sum_{l=1}^K \mathbbm{1}\left(\left(\tilde{x}_{j,k}  + y^{(k)}_k\right) \geq \left(\tilde{x}_{j,l} + y^{(k)}_l\right) \right)
    \frac{x_{j,l}+y^{(k)}_l}{N_j+1}\right]>\alpha\Big\}.\nonumber
\end{align}

By Theorem \ref{allconformal}, $\optpredsetj$ is an $\alpha$-valid procedure, and
it is constructed based on prior information.
Specifically, it differs from the direct set given in Equation \ref{directpredeqn} in that
categories are admitted into the prediction set based on an ordering determined by posterior counts that incorporate indirect information $\gammabf$, as opposed to an ordering based on the observed sample.
Moreover, it has been shown that if the indirect information used is accurate, $\optpredsetj$ may be more precise than a direct prediction set with the same coverage rate \citep{Hoff2023,Bersson2022}.


In total, $\directpredsetj$ and $\optpredsetj$ are both $\alpha$-valid prediction procedures.
They differ in the order in which species are admitted into the prediction sets, 
as species are admitted into the direct set in terms of decreasing empirical proportions and into the indirect set in terms of decreasing posterior counts. 
As a result, 
for an area with a small sample size, incorporating accurate prior information can 
result in an ordering used to construct a prediction set
that more accurately approximates the oracle ordering 
as the empirical proportions might be too unstable.
Of note, these two approaches are equivalent for a uniform prior $\gammabf = c\mathbf{1}$, for any constant $c$.
This includes, for example, a standard noninformative prior $c=1$, a standard objective Bayes Jeffrey's prior $c=1/2$, and an improper prior $c=0$.

\subsection{Empirical Bayes estimation of indirect information}\label{empBayesproc}

To obtain an $\alpha$-valid indirect prediction set for county $j\in\{1,...,J\}$, 
all that is required is an estimate of the prior concentration parameter $\gammabf$. 
We propose an empirical Bayesian approach whereby values of $\gammabf$ to be used for county $j$ are estimated from data collected in neighboring counties.
Specifically, we use the maximum likelihood estimate of the marginal likelihood based on the conjugate hierarchical model given by Equations \ref{samplingmodel} and \ref{linkingmodel},
\begin{align}\label{mll}
    \gammabf_j ={}& \arg\max_{\gammabf}\log p\left(\bigcup_{l\in L}\Xbf_l\Big|\gammabf\right)\\ 
    ={}& \arg\max_{\gammabf}\log \prod_{l\in L} \left[
    \frac{\Gamma(\sum_{i=1}^K \gamma_i)} {\Gamma(\sum_{i=1}^K x_{l,i}+\gamma_i)} \times\prod_{i=1}^K 
    \frac{\Gamma(x_{l,i}+\gamma_i)}{\Gamma(\gamma_i)}\right],\nonumber
\end{align}
where $L_j\subseteq \{1,...,K\}\backslash \{j\}$ is a non-empty set containing the indices of counties neighboring county $j$.
Information is shared across neighboring counties to inform an estimate of the prior for county $j$, and,
when estimated in this way, the prior concentration represents an across-county pooled prior concentration.
This optimization problem can be solved numerically with a Newton-Raphson algorithm.
See Appendix \ref{NRalgo} for details and derivation of such an algorithm.
Code to implement this procedure in the R Statistical Programming language is available online, see Section \ref{sec5}.

When $\gammabf_j$ is estimated using data independent of area $j$ and used to construct $\optpredsetj$,
the finite sample coverage guarantee of $\optpredsetj$ holds regardless of the accuracy of the estimated prior hyperparameter. 
If the estimated vector $\gammabf_j$ is accurate, then $\optpredsetj$ may also be 
more precise than direct prediction approaches. 

\section{Simulation Study}\label{sec3}

\begin{figure*}[t!]
\centering
{\includegraphics[width=.32\textwidth,keepaspectratio]{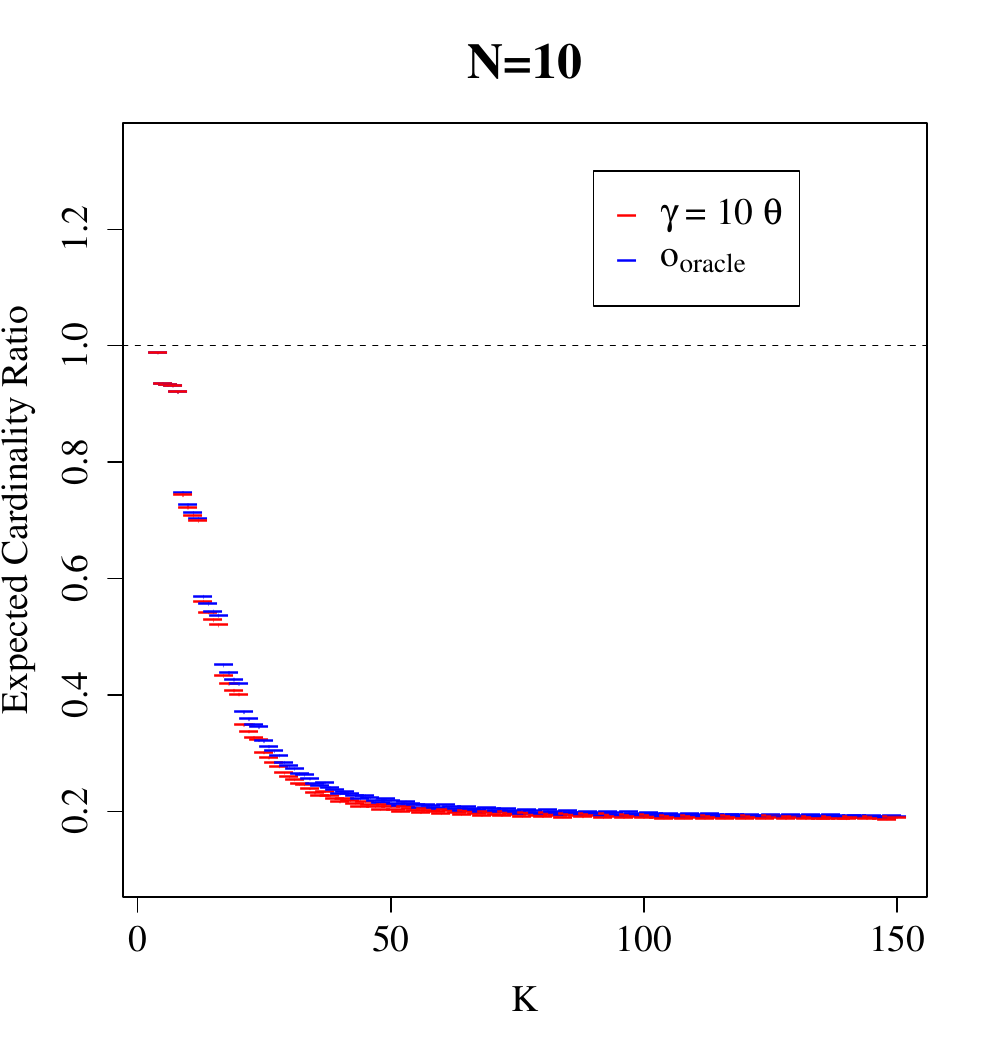}
\includegraphics[width=.32\textwidth,keepaspectratio]{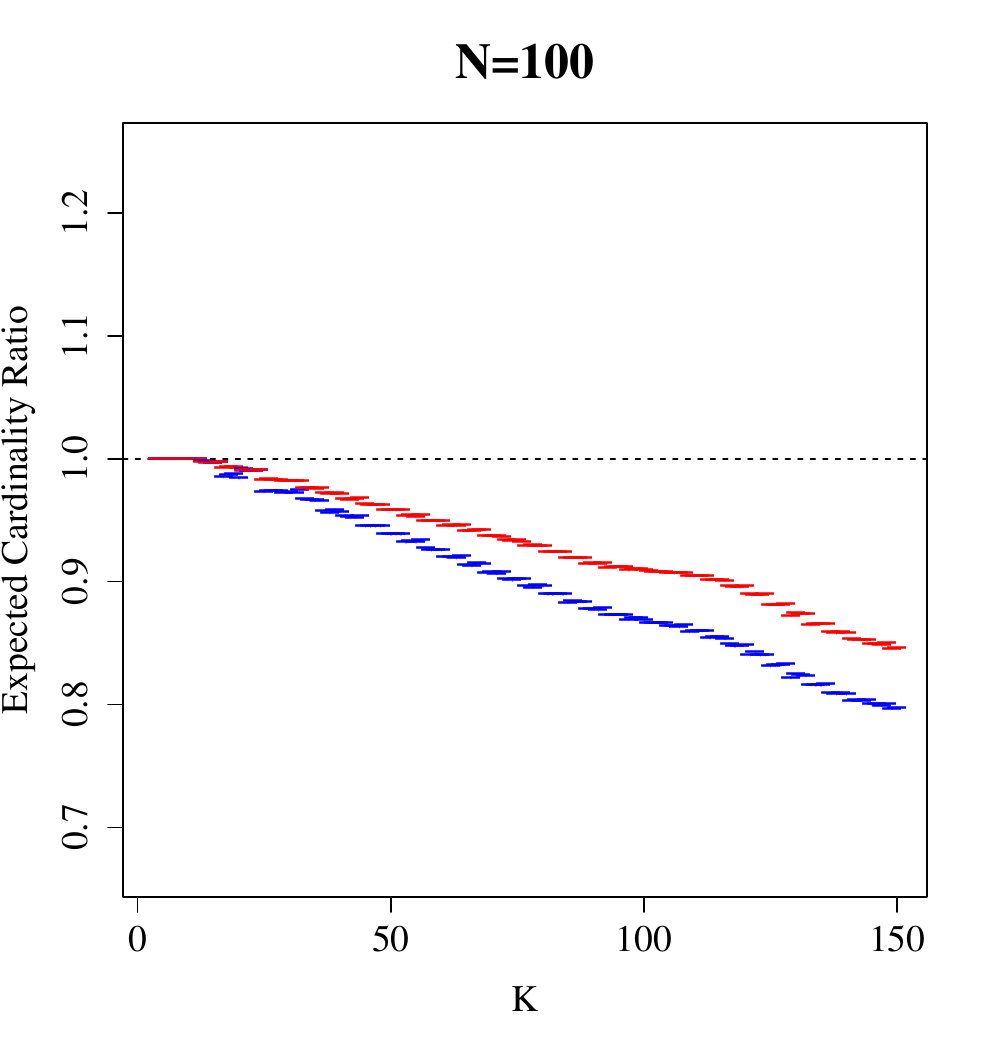}
\includegraphics[width=.32\textwidth,keepaspectratio]{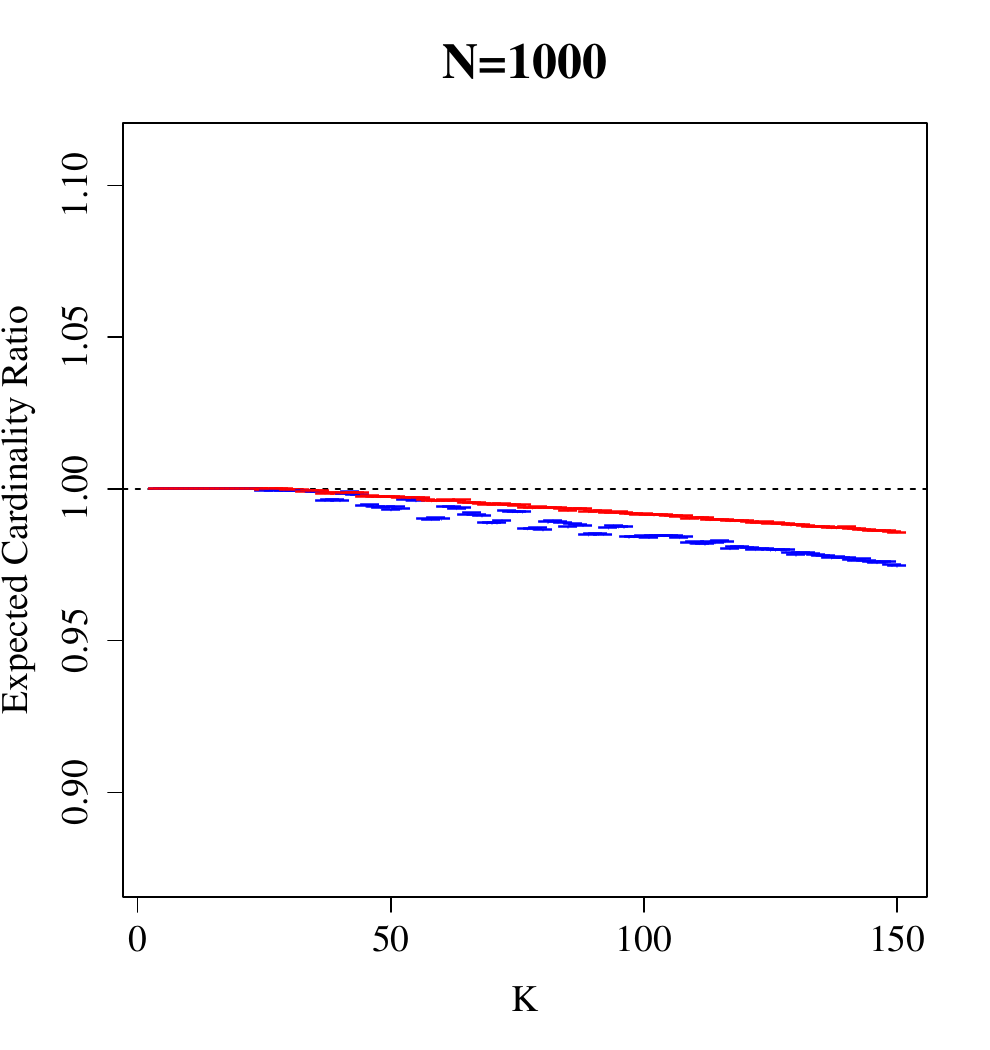}}
\caption{Monte Carlo approximations (+/- 1 standard deviation) of the expected cardinality ratio  of (red) indirect to direct methods and (blue) $\alpha$-valid prediction set given the oracle ordering to direct method. 
}\label{numeranal}
\end{figure*}

To illustrate how the incorporation of indirect information can affect precision of prediction sets,
we compare expected set cardinality 
obtained from the indirect and direct prediction methods for a single simulated area. 
In contrast to the eBird data, for example, the analysis of this section corresponds to that of one county.
Because citizen science data such as these often feature unequal sampling efforts across counties, we are particularly interested in demonstrating the difference in cardinality between these two approaches for a range of sample sizes $N=10,100,1000$.
Moreover, we compare results for varying number of categories $K$.
Throughout, we consider
a low entropy regime in which $\lceil K/4 \rceil$ categories unequally split nearly all of the probability mass,
and the rest of the categories have nearly probability $0$.
While we do not necessarily expect real populations in practice to have such a distribution, 
it is chosen to clearly demonstrate the benefit of
including indirect information in the construction of prediction sets that maintain frequentist coverage.

In one construction of indirect prediction sets, we consider a prior based on full information 
with moderate prior precision $\gammabf = \thetabf \times 10$.
We compare with direct prediction sets given by Equation \ref{directpredeqn}, or, equivalently, indirect prediction sets constructed with a uniform prior $\gammabf = c\mathbf{1}$.
Finally, we compare the approaches to $\alpha$-valid order-based prediction sets obtained based on an oracle ordering.
Results comparing Monte Carlo approximations of the expected prediction set cardinality ratios between the various approaches obtained from $25,000$ replications 
are displayed in Figure \ref{numeranal}. 

As all methods considered are $\alpha$-valid procedures, the crucial difference between them is the incorporation of indirect information. 
Utilizing accurate prior information in the construction of prediction sets generally results in prediction sets distinctly smaller
than direct sets, particularly so 
if there are a large number of categories relative to the sample size. 
This is evidenced by the red dashes in Figure \ref{numeranal} showing the expected cardinality ratios of the indirect to direct prediction sets are always at or below a value of 1.
An accurate prior may be one that 
approximates the true probability mass vector well
with large precision relative to sample size, as seen in the left plot of Figure \ref{numeranal} for sample size $N=10$.
More generally, though, all that is needed is 
a prior that results in 
posterior counts that
accurately approximate the oracle ordering of categories.
We discuss the three sample size regimes in detail below.

For a small sample size of $N=10$, the prior $\gammabf$ used to construct the indirect prediction sets is an informative prior with strong precision in that the scale used is equal to the sample size in this case. 
As a result, the posterior distributions contain notably more information than what is in each simulated dataset. 
As a result, the ordering of categories induced by the posterior counts, used to construct the indirect prediction sets, are accurately approximating the oracle ordering of categories. This is evidenced by the 
nearly identical behavior of the
two cardinality ratios explored.
In conjunction with the instability of the direct method in the presence of such a small sample size,
this results in notably smaller cardinality of the indirect set as compared to the direct set, even for relatively small total numbers of categories. 
At its best, the indirect prediction set is about 80\% smaller than the direct set.

For a moderate sample size of $N=100$,
the prior precision used to construct the indirect prediction sets is not overwhelming as compared to the sample size,
and hence the posterior counts do not approximate the oracle ordering as well as in the regime with a smaller sample size.
This is evidenced by the divergence of the red and blue dashes in the middle plot of Figure \ref{numeranal}.
Still, particularly as the number of categories increases for fixed $N$, the benefit of utilizing prior information of this type is highlighted by the decline of the cardinality ratio of the indirect to direct prediction sets (red lines).
For example, in the case of $N=100$ and $K=150$, the indirect prediction set constructed with $\gammabf$ is about $15\%$ smaller than the direct prediction set.

A similar but less pronounced pattern is seen in 
the presence of 
a larger sample size of $N=1000$. For this sample size with $K\leq 150$, all methods considered perform relatively similarly.
However, 
as the number of categories increases,
there is a distinct gain in prediction set precision given the input of indirect information in prediction set construction.

\section{Summarizing eBird species abundance data}\label{sec4}

\begin{figure*}[t!]
\centering
\includegraphics[width=.8\textwidth,keepaspectratio]{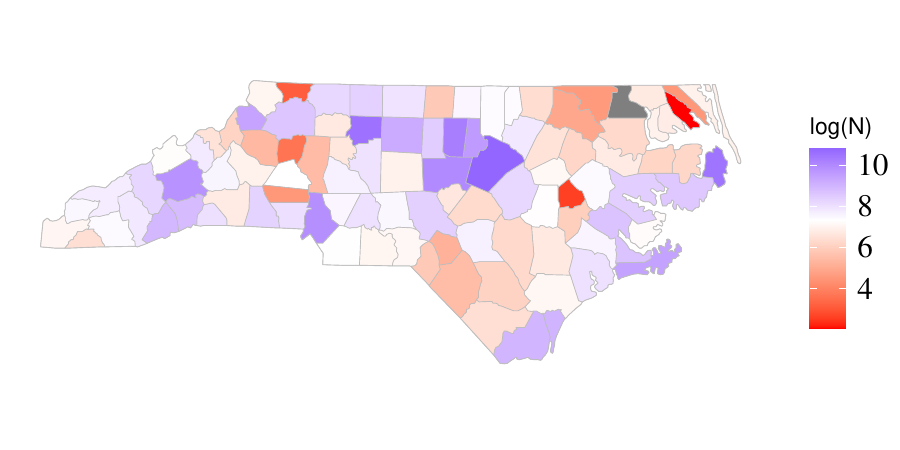}
\caption{Within-county log sample size.}\label{ebird_ss}
\end{figure*}

In this section, we 
describe avian species abundance in North Carolina, USA 
from eBird data obtained from citizen-uploaded complete checklists of species observations
in the first week of May 2023. 
Across the $99$ counties, 
$393$ unique species were identified. 
Some species such as the Northern Cardinal, Carolina Wren, and American Robin were identified frequently. 
Many others like the Northern Saw-whet Owl and the Solitary Sandpiper were rarely seen; in fact, $50\%$ of species were seen fewer than $100$
times each across the entire state.
Moreover, within-county sample sizes vary drastically (Figure \ref{ebird_ss}) from approximately $50,000$ individual birds identified in Wake County, one of the most populous counties in NC that contains the state's capital, to only $8$ in Pasquotank County, a small coastal county consisting of about $1/30$\textsuperscript{th} of the human population of Wake County.  

As motivated in the Introduction, describing such data with $\alpha$-valid prediction sets for each county provides a useful summary with unambiguous statistical interpretation. 
That is, with at least probability $1-\alpha$, 
an unobserved bird in a given county will belong to a species contained in the specified prediction set,
where the probability is taken with respect to the random sample and the predictand. 
Here, we demonstrate the usefulness of this approach 
in gaining better understanding of species abundance. 
Moreover, we elaborate on the benefit of utilizing indirect information in 
the construction of practically useful sets that are precise, 
particularly for counties with small within-county sample sizes.

\begin{figure*}[t!]
\centering
\includegraphics[width=.8\textwidth,keepaspectratio]{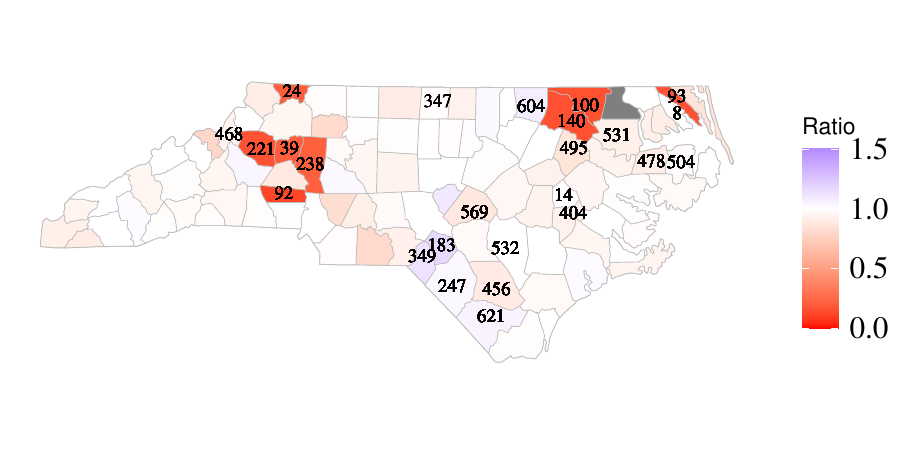}
\caption{Cardinality ratio of indirect to direct prediction sets. Prior hyperparameters estimated with an empirical Bayesian procedure based on five nearest neighbors for each county. The lowest quantile sample sizes are overlaid on their respective counties.}\label{ebird_ratio}
\end{figure*}

For each county in NC, we construct an indirect prediction set 
based on a prior hyperparameter estimated from data in the five nearest neighboring counties,
following the procedure described in Section \ref{empBayesproc}.
The eBird data consist of independent samples collected across the state, so samples are independent across counties. As a result of this independence,
finite-sample coverage of the indirect prediction approach is guaranteed. 
We compare the cardinality of these indirect prediction sets to that of direct prediction sets, both of which maintain at least $95\%$ coverage for each county. 
The cardinality ratios of the indirect to direct prediction sets across the counties in NC are plotted in Figure \ref{ebird_ratio}.
To highlight the impact of within-county sample size, the lower quantile sample sizes are overlaid on their respective county.

In general, the incorporation of indirect information in the construction of prediction sets results in notably smaller cardinality of the indirect prediction sets as compared with that of the direct prediction sets. 
Of the 99 counties in NC, indirect sets have smaller cardinality in 65, and the two approaches result in the same cardinality in 20 counties. 
The improvement in cardinality is particularly conspicuous in counties with small to moderate sample sizes, as evidenced by the sample sizes of counties with the brightest shade of red in Figure \ref{ebird_ratio}.
Moreover, ten counties have trivial direct sets consisting of all $K$ species,
while only two counties with the smallest within-county sample sizes, 8 and 14, have trivial indirect prediction sets.
For the county with the third smallest sample size (24), the indirect prediction set
only includes 80 species, or about 20\% of all possible species, while the direct prediction set is the trivial set.

Overall, even in counties with larger sample sizes, it is most common for the indirect and direct prediction sets to 
contain a different set of species.
In fact,
the indirect and direct prediction sets disagree for nearly every county in NC. They are equivalent for only six counties where they aren't both trivial sets.
Commonly, 
this discrepancy corresponds with smaller indirect sets, and hence highlights the benefit of inclusion of indirect information in the construction of prediction sets.

\begin{figure*}[htb]
\centering
\includegraphics[width=.95\textwidth, keepaspectratio]{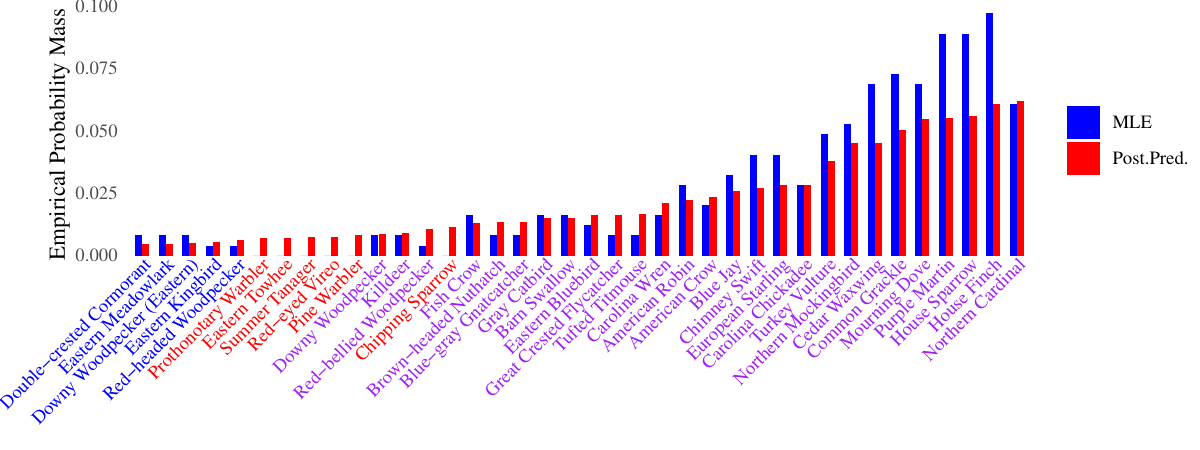}
\caption{Empirical probability masses of species included in only the indirect (red text), only the direct (blue text), or both (purple text) prediction sets, sorted by the posterior proportion. 
MLE plotted in blue, and the posterior proportion based on a prior estimated from data in nearest five neighboring counties.}\label{NHC_mass}
\end{figure*}

\subsection{Order-based prediction in Robeson County}


\begin{table*}[b!]
\centering
\begin{tabular}{rllll}
  \hline
 & \blue{D. Cormorant} & \blue{E. Kingbird} & \red{Pine Warbler} & \red{C. Sparrow} \\ 
  \hline
\textbf{Robeson} & 0.81\% & 0.4\% & 0.00\% & 0.00\% \\ 
   NC-017 & 0.00\% & 2.85\% & 1.97\% & 2.63\% \\ 
   NC-047 & 0.00\% & 0.00\% & 1.29\% & 0.64\% \\ 
   NC-051 & 0.00\% & 0.68\% & 2.62\% & 5.59\% \\ 
   NC-093 & 0.00\% & 0.00\% & 1.09\% & 0.55\% \\ 
   NC-165 & 0.00\% & 0.86\% & 2.58\% & 5.16\% \\
  \hline \hline
  $\gammabf$ & 0.00 & 1.33 & 3.42 & 4.69   \\ 
   \hline
\end{tabular}
\caption{Percentage of all birds observed within each respective county,
for select species included in either the indirect set (red text)
or the direct set (blue text). Estimated prior hyperparameter $\gammabf$ for Robeson County recorded in the last row.}\label{neighborperc}
\end{table*}

To further compare the two approaches and elucidate the role of the ordering of the species, we elaborate on the construction of indirect and direct prediction sets for Robeson County.
Robeson is located near the southeastern border of NC and features a moderately small within-county sample size of $247$ birds observed,
with species-specific observation counts ranging from zero to ten.
The two prediction sets have nearly the same cardinality but contain differences in species inclusion.
Specifically, the indirect prediction set contains $33$ species, and the direct set contains $32$, with an overlap of $27$ species.

To illustrate the role of the ordering used in the construction of $\alpha$-valid prediction sets, 
the empirical proportions based on the observed sample (MLE) and posterior proportions (Post.Pred) are plotted in Figure \ref{NHC_mass} for the union of included species in the two sets.
In the figure, the species are sorted by increasing posterior proportions. 
The indirect and direct sets include species based on the posterior and empirical distributions, respectively.
Discrepancies between the indirect and direct sets occur when these two distributions disagree.
From Figure \ref{NHC_mass}, it is easy to see the indirect prediction set consists of the species with the 33 largest posterior predictive proportions. 
In contrast, the direct set consists of species with the largest sample probability mass.
Naturally, the ordering of these two estimates agree for species common to the region, and, as such, there is a fair amount of overlap of species inclusion.

As a result of our estimation procedure for the prior hyperparameter $\gammabf$ for Robeson County, the disparity between inclusion or exclusion of a species among the two prediction set methods
is further elucidated by examining species presence in neighboring counties.
In short, species with more frequent occurrence in neighboring counties will have a larger estimated prior count than those seen rarely in neighboring counties.
Species occurrences in neighboring counties are displayed in Table \ref{neighborperc} for a select few species along with the estimated $\gammabf$ for Robeson County, obtained by solving Equation \ref{mll} using data in these neighboring counties.

Intuitively, species
that are seen
in neighboring counties with some relative frequency, such as the Chipping Sparrow or Pine Warbler, 
are probably also present in Robeson County, and hence
should be included in a prediction set.
In practice, these species
have a comparatively high estimated prior of about 5 and 4, respectively, and hence are included in the indirect prediction set even though they weren't recorded as being observed in Robeson County in the dataset.
Alternatively,
consideration of indirect information yields the conclusion that 
species like the Eastern Kingbird and Cormorant may be rare in the area in general, as reflected by small $\gammabf$ values,
and thus these species are not included in the indirect prediction set.

\subsection{Inference among species with tied observed counts in Haywood County}

\begin{table*}[ht]
\centering
\begin{tabular}{rlllll}
  \hline
 & \blue{L. Flycatcher} & \blue{R. Hawk} & \blue{C. Yellowthroat} & \blue{E. Kingbird} & \purple{Bobolink}  \\ 
  \hline
\textbf{Haywood} & 0.21\% & 0.24\% & 0.21\% & 0.24\% & 0.24\% \\ 
   NC-021 & 0.06\% & 0.29\% & 0.18\% & 0.43\% & 0.14\%   \\ 
   NC-099 & 0.00\% & 0.08\% & 0.16\% & 0.00\% & 0.00\%  \\ 
   NC-115 & 0.07\% & 0.14\% & 0.21\% & 0.28\% & 0.00\%   \\ 
   NC-173 & 0.18\% & 0.18\% & 0.66\% & 0.09\% & 1.41\%   \\ 
   NC-175 & 0.00\% & 0.30\% & 1.00\% & 0.54\% & 0.84\%  \\ 
   \hline \hline
  $\gammabf$ & 0.4 & 1.29 & 2.43 & 1.49 & 2.15  \\ 
   \hline
\end{tabular}
\caption{Percentage of all birds observed within each respective county for species included in either both prediction sets (purple text) or only the direct set (blue text).
Estimated prior hyperparameter $\gammabf$ for Haywood County recorded in the last row.
Species are sorted by posterior proportion. }\label{Haywood_table}
\end{table*}

In
species abundance data, particularly for areas or counties with small sample sizes, it is common for multiple species to have the same observed count.
A feature of the construction of the direct order-based prediction approach as presented is that species with the same observed counts
will either be jointly included or excluded from the prediction set.
As a result, 
a direct prediction set constructed
from a sample with tied species counts 
may have increased cardinality over an indirect prediction set that does not necessarily jointly admit all species with tied observed counts. 
If the direct set has increased cardinality for this reason, the direct set will also have increased coverage over the indirect set.

When constructing a prediction set based on the empirical proportions without consideration of indirect information, as in the construction of the direct set, 
this may commonly occur, and 
there is 
no clear approach
to choose among the species with tied counts without further information than what is provided in the sample in that county. 
One could randomly choose to include one of the species from the set of species with tied counts, for example, but 
a more principled manner 
is to utilize indirect information to determine 
which species should be included.
This is the mechanism used by the indirect prediction approach when the prior hyperparameter is a real valued vector estimated from indirect information. 
As such, a more nuanced benefit of utilizing indirect information in the construction of a prediction set is the capacity to include a select few categories with tied empirical proportions.

To demonstrate, we elaborate on species inclusion in the indirect and direct prediction sets in Haywood County.
Haywood is popular destination in the Blue Ridge Mountains, located near the western border of North Carolina.
It features a moderately large within-county sample size of roughly $4000$ birds observed.
In Haywood County, the indirect prediction set contains $70$ species, and the larger direct set contains $74$. 
In the construction of these prediction sets, the ordering of species with regards to the posterior proportions and the empirical proportions agree for most species. 
As a result, all $70$ species included in the indirect set are also included in the direct set. 
The disparity in species inclusion occurs primarily as a result of 
tied counts of species occurrence in the sample.

Empirical proportions in Haywood and neighboring counties are reported in Table \ref{Haywood_table}
for the five species included in Haywood County's prediction sets with the smallest posterior proportions.
The species with the four smallest posterior proportions are included only in the direct set, and the other species, the Bobolink, is included in both the indirect and direct sets.
The Bobolink was observed $9$ times in the sample from Haywood County, or about $0.24\%$ of the Haywood sample. 
For an ordering determined by either the empirical counts or the posterior counts, this species is required to be included in the order-based prediction set to guarantee $1-\alpha$ coverage.
Two of the other species, the Red-shouldered Hawk and Eastern Kingbird, were each also observed $9$ times in the sample from Haywood, and, by construction of the order-based prediction approach, must also be included in the direct set.
When admitting the species into a prediction set by posterior counts based on the real-valued prior hyperparameter $\gammabf$ estimated from data in neighboring counties, as in the indirect approach considered, the `tie' among these three species is broken, and only one, the Bobolink, is included in the indirect prediction set.


\section{Discussion}\label{sec5}


Species abundance data collected across heterogeneous areas is increasingly important in understanding biodiversity. 
Some of the largest sources of such data are citizen science databases for which volunteers spearhead the data collection.
As a result of the civilian-led scientific effort, such data often feature unequal sampling across a spatial domain where some areas have large within-area sample sizes and others have much smaller within-area sample sizes.

In this article, we propose summarizing species abundance data of this type with valid prediction sets that are constructed by sharing information across areas.
Utilizing indirect information may result in smaller prediction sets than otherwise achievable with direct methods.
Meanwhile, maintaining validity of the prediction sets for each area allows for an accessible interpretation that enables a straightforward comparison across areas.
In particular, maintaining interpretable statistical guarantees on a descriptor of such data is important as analyses from such data often have far reaching policy implications.
Smaller prediction sets may be attainable based on Bayesian inference of a spatial hierarchical model such as that presented in \cite{Tang2023}, for example,
but these approaches introduce bias and 
a resulting prediction set would not retain the nominal frequentist coverage rate guarantee for each county.

The usefulness of our approach for 
summarizing citizen science data
is motivated in part to combat the common problem of varying sampling efforts across areas.
We detail how $\alpha$-valid prediction sets can be constructed 
with the incorporation of indirect information to 
improve within-county prediction set precision and
propose an empirical Bayes procedure to do so.
Incorporation of accurate indirect information
results in a narrower prediction set for a given county than a direct prediction set
by exploiting data in nearest neighboring counties.
The proposed empirical Bayes procedure is based on a standard hierarchical model that is straightforward to understand, and the authors provide code for implementation.

There may, however, be a benefit to utilizing a more structured prior
that incorporates indirect information in a more complex manner such as a prior that weights data from different parts of the state differently. 
For example, a model based on a learned intrinsic distance between counties was shown in \cite{Christensen2022}
to fit a subset of the eBird data better than standard methods based on geographic adjacency structure.
In the sample analyzed in Section \ref{sec4}, we found an indirect prediction set constructed with a hyperparameter estimated from five nearest neighbors results in overall narrower prediction sets than a direct approach, but 
it would be valuable to explore if
this can be further improved upon with a more detailed prior. 
More broadly, different applications may warrant an alternative information sharing prior if, for example, there is no notion of spatial distance across the different areas. 
For example, it may be of interest to compare species abundance variation across different time frames for a given county.

All replication codes for this article, including functions to implement the empirical Bayes estimation procedure for the prior hyperparameter, are available at \url{https://https://github.com/betsybersson/FreqPredSets_Indirect}.

\bibliographystyle{misc/spbasic}
\bibliography{misc/library}

\appendix

\section{Maximization of the marginal multinomial-Dirichlet likelihood}\label{NRalgo}

In this section, we detail a Newton-Raphson algorithm to maximize the marginal log likelihood of a conjugate multinomial-Dirichlet model:
\begin{align*}\label{mll_eqn}
    \Xbf_j\sim{}& MN_K(\thetabf_j,N_j),\;\text{independently for }j=1,...,J\\
    \thetabf_1,...,\thetabf_J\sim{}&Dirichlet_K(\gammabf).\nonumber
\end{align*}
The log likelihood of the marginal likelihood is as follows,
\begin{align*}
\mathcal{L}(\gammabf){}& \propto \sum_{j=1}^J \Bigg[ 
\log \Gamma(\sum_{i=1}^K \gamma_i) -\log\Gamma(N_j+\sum_{i=1}^K \gamma_i) + \\
{}& \sum_{i=1}^K\log\Gamma(x_{j,i}+\gamma_i) - \sum_{i=1}^K\log\Gamma(\gamma_i) \Bigg].
\end{align*}
Define
\[
\Psi(s) = \frac{d}{ds}log \Gamma(s) = -\xi +\sum_{n=0}^\infty\left[\frac{1}{n+1}-\frac{1}{n+s}\right],
\]
where $\xi$ is the Euler-Mascheroni constant.
Then, it is straightforward to obtain the first and second derivatives of the marginal log likelihood, 
\begin{align*}
\frac{d}{d\gamma_k} ={}& \sum_{j=1}^J\Bigg[ \Psi(\sum_{i=1}^K\gamma_i) -\Psi(N_j+\sum_{i=1}^K\gamma_k) +\Psi(x_{j,k}+\gamma_k)- \\
{}& \hspace{1.5cm} \Psi(\gamma_k)\Bigg]\nonumber \\
\frac{d}{d\gamma_k^2} ={}& \sum_{j=1}^J\Bigg[ \Psi'(\sum_{i=1}^K\gamma_i) -\Psi'(N_j+\sum_i\gamma_i) + \Psi'(x_{j,k}+\gamma_k)-\\
{}& \hspace{1.5cm}\Psi'(\gamma_k)\Bigg]\nonumber\\
\frac{d}{d\gamma_kd\gamma_{k'}} ={}& \sum_{j=1}^J\left[ \Psi'(\sum_{i=1}^K\gamma_i) -\Psi'(N_j+\sum_{i=1}^K\gamma_i) \right],
\end{align*}
where $\Psi'$ is the trigamma function. 
Let $\boldsymbol{g}$ be the gradient vector of length $K$ 
and $\boldsymbol{H}$ the Hessian matrix. 
Finally, Newton's method updates $\gammabf$ as follows:
\[
\gammabf^{(t+1)} = \gammabf^{(t)}- \boldsymbol{H}^{-1}(\gammabf^{(t)})\boldsymbol{g}(  \gammabf^{(t)}),
\]
where the algorithm is iterated until convergence.

\end{document}